\documentclass[%
 aip,
 jcp,%
 amsmath,amssymb,
 preprint,
]{revtex4-2}

\usepackage{graphicx}
\usepackage{dcolumn}
\usepackage{bm}
\usepackage[usenames,dvipsnames]{color}

\usepackage{cancel}
\usepackage[dvipsnames]{xcolor}

\usepackage[normalem]{ulem}

\usepackage{tabularx, threeparttable}
\newcolumntype{L}[1]{>{\raggedright\arraybackslash}p{#1}} 
\newcolumntype{C}[1]{>{\centering\arraybackslash}p{#1}} 
\newcolumntype{R}[1]{>{\raggedleft\arraybackslash}p{#1}} 

\begin{document}

\setlength\parindent{0pt}


\title[]{Letter: GROMACS Stochastic Dynamics and BAOAB are equivalent configurational sampling algorithms}%

\author{Stefanie Kieninger}
\affiliation{Department of Biology, Chemistry, Pharmacy, Freie Universit\"{a}t Berlin, Arnimallee 22, D-14195 Berlin, Germany}

\author{Bettina G. Keller}
\email[]{bettina.keller@fu-berlin.de}
\affiliation{Department of Biology, Chemistry, Pharmacy, Freie Universit\"{a}t Berlin, Arnimallee 22, D-14195 Berlin, Germany}


%
%

%
%
\begin{abstract}
Two of the most widely used Langevin integrators for molecular dynamics simulations are the GROMACS Stochastic Dynamics (GSD) integrator and the splitting method BAOAB. In this letter, we show that the GROMACS Stochastic Dynamics integrator is equal to the less frequently used splitting method BAOA. It immediately follows that GSD and BAOAB sample the same configurations and have the same high configurational accuracy.  Our numerical results indicate that GSD/BAOA has higher kinetic accuracy than BAOAB. 
\end{abstract}

\maketitle


\section{Introduction}

Langevin integrators, or equivalently Langevin thermostats, are widely used in molecular dynamics (MD) simulations to ensure that the simulation correctly samples the canonical ensemble \cite{Huenenberger:2005}.
Two of the most widely used Langevin integrators are the GROMACS Stochastic Dynamics integrator (GSD)\cite{Goga:2012} and the BAOAB integrator \cite{Leimkuhler:2012, Leimkuhler:2013, Bussi:2007}.
The GSD integrator is derived by extending the leap frog algorithm for deterministic dynamics
by an impulsive application of friction. 
The algorithm exhibits first-order kinetics in the temperature relaxation and faithfully reproduces diffusion constants for a wide range of collision rates \cite{Goga:2012}. 
The algorithm has been implemented as the standard Langevin integrator in GROMACS \cite{Abraham:2015} and is widely used in 
atomistic simulations \cite{Thurston:2016,Aldeghi:2017, Mansbach:2021}, 
coarse-grained simulations \cite{Marrink:2013, Goga:2015, Deichmann:2018, Pezeshkian:2020}, and 
dissipative particle simulations \cite{Goicochea:2015, Moga:2013}.
Besides its application within GROMACS, GSD has also been implemented and used  in multiple other studies \cite{Schneider:2017, Jung:2017, Moga:2013, Moga:2015, Madeo:2020}.

The BAOAB integrator is based on splitting  the vector field in the Langevin equation of motion into parts labelled A, B and O, and integrating these parts separately.
This yields update operators $\mathcal{A}$, $\mathcal{B}$, and $\mathcal{O}$, 
where $\mathcal{A}$ represents a deterministic step in the configuration space, $\mathcal{B}$ a deterministic step in the momentum space, and $\mathcal{O}$ represents the update in momentum space due to the friction force and the random force. (See supplementary material).
The name BAOAB encodes the sequence of the operators, where update operators that appear twice are carried out for half a time step.
BAOAB has been implemented in OpenMMTools
\cite{openmmtools_github, Fass:2018}
for the MD package OpenMM \cite{Eastman2017} and is used frequently in atomistic MD simulations 
\cite{Brotzakis:2019, Starr:2021}.
The method has been a starting point for the development of integration schemes for dynamics beyond classical MD, such as
nonequilibrium MD \cite{Sivak:2013},
generalized Langevin dynamics \cite{Baczewski:2013, Ple:2019, Duong:2021}, 
ab-initio path-integral MD \cite{Liu:2016, Lopanitsyna:2021}, and 
multi-timestep MD \cite{Leimkuhler:2013b, Lagardere:2019}. 
For any Langevin integrator, a systematic error in the sampled configurational Boltzmann density arises with increasing timestep. 
This error has been shown (analytically and numerically) to  be particularly small in BAOAB \cite{Leimkuhler:2012, Leimkuhler:2013, Sivak:2013, Sivak:2014, Fass:2018}.
Thus, BAOAB can be operated at large timesteps - at least if the goal of the simulation is to sample the configurational Boltzmann density.
For simulations of water, numerical stability up to timesteps of 6 to 9 fs have been reported \cite{Fass:2018, Leimkuhler:2016}.
%

%
BAOAB is closely related to the BAOA integrator \cite{BouRabee:2010}, as has previously been noted \cite{Leimkuhler:2012, Zhang:2019, Song:2021}:
they sample the same positions and only differ by a shift of $\frac{\Delta t}{2}$ in the momenta.
BAOA has recently been implemented in the MD packages AMBER \cite{Case:2021} and OpenMM  \cite{Eastman2017}, where the algorithm is called  LFMiddle \cite{Zhang:2019}.
These three algorithms for MD simulations have so far been treated as separate integrators in the literature.
Here, we argue that GSD and BAOAB sample the same configurations and similar momenta.
For this, we show that the GSD equations can be rearranged to yield the BAOA integrator \cite{BouRabee:2010}.
We show that this similarity also extends to GSD.

%

\section{Theory}
\subsection{Langevin dynamics}

Consider a particle with mass $m$ that moves in a one-dimensional position space $q \in \mathbb{R}$ according to underdamped Langevin dynamics
\begin{align}
    \dot{q} &= \frac{p}{m}   \cr
    \dot{p} &= - \nabla_q V(q) - \xi p + \sqrt{2\xi k_BTm} \, \eta(t) \, ,
 \label{eq:underdamped_Langevin_01}    
\end{align}
where $V(q)$ is the potential energy function at position $q$, $\nabla_q = \partial/\partial q$ denotes the gradient with respect to the position coordinate,
$\xi$  is a collision or friction rate (in units of s$^{-1}$),
$T$ is the temperature and $k_B$ is the  Boltzmann constant. 
$\eta \in \mathbb{R}$ is an uncorrelated Gaussian white noise with unit variance centered at zero
$ \langle \eta(t)\eta(t') \rangle = \delta(t-t')$,
where $\delta(t-t')$ is the Dirac delta-function.
We use the dot-notation for derivatives with respect to time: $\dot q = \partial q / \partial t$.
$\omega(t) = (q(t), p(t)) \in \Omega \subset \mathbb{R}^2$ denotes the state of the system at time $t$, which consists of positions $q(t)$ and conjugated momenta $p(t) = m\dot{q}(t)$.
$\Omega$ is called state space or phase space of the system. \\

A Langevin integrator $I$ is a numerical integration scheme that solves eq.~\ref{eq:underdamped_Langevin_01}. 
It produces a time-discretized approximation $\boldsymbol{\omega}^I$ of a continuous trajectory $\omega(t)$ 
\begin{eqnarray}
    \omega(t) \approx  \boldsymbol{\omega}^I = (\omega_1^I, \omega_2^I \dots, \omega_n^I \,|\, \omega_0)  \,  ,
\end{eqnarray}
where $\omega(t)$ denote an exact solution of eq.~\ref{eq:underdamped_Langevin_01}.
The notation emphasises that the programme needs the initial state of the system $\omega_0=(q_0,p_0)$ as an input (initial condition).
$n$ is the number of integration timesteps $\Delta t$. 
We will denote the index of the integration timestep by $k$, such that $\omega(t = k \Delta t) \approx \omega_k^I$, and analogously 
$q(t = k \Delta t) \approx q_k^I$ and
$p(t = k \Delta t) \approx p_k^I$.\\
Each of the Langevin integrators discussed here uses a single random number $\eta_k$ per integration step (and degree of freedom).
Thus, given an initial state $\omega_0$, the timestep $\Delta t$, a potential energy function $V(q)$ and a random number sequence $\boldsymbol{\eta}=(\eta_1,\eta_2,\dots,\eta_n)$, the trajectory $\boldsymbol{\omega}^I$ is defined unambiguously for a Langevin integrator $I$.
However, with the same parameters different Langevin integrators yield slightly different trajectories, e.g. $\boldsymbol{\omega}^{\mathrm{BAOAB}} \ne \boldsymbol{\omega}^{\mathrm{ABOBA}}$.
These differences determine the different numerical accuracies of  Langevin integrators.

\subsection{GSD equals BAOA\label{sec:GSD=BAOA}}

We show that the GSD integrator is equal to a BAOA splitting algorithm.
In Ref.~\onlinecite{Goga:2012}, the GSD integrator is reported with the following equations
\begin{subequations}
\begin{align}
    p           &= p_{k- \frac{1}{2}}  -\nabla V(q_k) \Delta t  \label{eq:Goga01} \\
    \Delta p    &= -f p  + \sqrt{f(2-f) m k_BT} \eta_k \label{eq:Goga02} \\
    q_{k+1}     &= q_k +\left(\frac{p}{m} + \frac{1}{2} \frac{\Delta p}{m}\right) \Delta t \label{eq:Goga03} \\
    p_{k + \frac{1}{2}} &= p + \Delta p\label{eq:Goga04} \, ,
\end{align}
\end{subequations}
where we made the following changes in the notation to be consistent with the notation in this contribution:
$h \rightarrow \Delta t$,
$a = F(t)/m \rightarrow -\frac{\nabla V(x(t))}{m}$, and
$\xi \rightarrow \eta$.
We also denoted positions by $q$ instead of $x$, and converted velocities into momenta: $v = p/m$.
Finally, we multiplied eqs.~\ref{eq:Goga01}, \ref{eq:Goga02} and \ref{eq:Goga04} by $m$ to obtain the algorithm in terms of the momenta.
We changed time $t$ into an iteration index $k$.
At the beginning of the integration step, position $q_k$, energy gradient $-\nabla V(q_k)$, and momentum $p_{k-\frac{1}{2}}$ are known. 
Because the algorithm is an extension of the deterministic leap frog algorithm, the momenta generated in each iteration of the integrator are assigned to half timesteps $k+\frac{1}{2}$.
In Ref.~\onlinecite{Goga:2012} the factor $f$ is initially a user-defined parameter, but is later related to the friction coefficient $\xi$ by
\begin{eqnarray}
    \xi = -\frac{1}{\Delta t} \ln (1-f) 
    &\Leftrightarrow&
    f  = 1 - e^{-\xi \Delta t }  
\label{eq:Goga_f1}    
\end{eqnarray}
(eq.~19 in Ref.~\onlinecite{Goga:2012}), and thus
\begin{eqnarray}
    f(2-f) 
    &=& \left(1-e^{-\xi \Delta t }\right) \cdot \left(1+e^{-\xi \Delta t }\right) 
    =   \left(1-e^{-2\xi \Delta t }\right) \, .
\label{eq:Goga_f2}        
\end{eqnarray}
Inserting eqs.~\ref{eq:Goga_f1} and \ref{eq:Goga_f2} into eq.~\ref{eq:Goga02} yields 
\begin{subequations}
\begin{align}
    \Delta p    &= - p + e^{-\xi \Delta t} p  + \sqrt{\left(1-e^{-2\xi \Delta t }\right)  m k_BT} \eta_k \, . \label{eq:Goga06} 
\end{align}
\end{subequations}
Then eq.~\ref{eq:Goga04} reduces to
\begin{eqnarray}
    p_{k + \frac{1}{2}} &= + e^{-\xi \Delta t} p  + \sqrt{\left(1-e^{-2\xi \Delta t }\right)  m k_BT} \eta_k \, . \label{eq:Goga_momentumUpdate}
\end{eqnarray}
That is, eq.~\ref{eq:Goga04} can be expressed in terms of only $p$, and the intermediate calculation of $\Delta p$ in eq.~\ref{eq:Goga06} can be omitted for the update of the momenta.
The position update in eq.~\ref{eq:Goga03} is a combination of two half-steps \cite{Goga:2012}: one after the first update of the momenta, $p_{k-\frac{1}{2}} \rightarrow p$, and one after the second update of the momenta $p \rightarrow p_{k+\frac{1}{2}}$:
\begin{align}
    q_{k+1}     
    &= q_k +\left(\frac{p}{m} + \frac{1}{2} \frac{\Delta p}{m}\right) \Delta t \cr
    &= q_k +\frac{p}{m} \frac{\Delta t}{2} + \left(\frac{p}{m}+ \frac{\Delta p}{m}\right)\frac{\Delta t}{2} \, ,
\end{align}
where the equations for the two half-steps are
\begin{subequations}
\begin{eqnarray}
    q_{k+\frac{1}{2}} &=& q_k + p\frac{\Delta t}{2m} \label{eq:Goga_positionUpdate01} \\
    q_{k+1} &=& q_{k+\frac{1}{2}} + p_{k+\frac{1}{2}}\frac{\Delta t}{2m} \,.
\label{eq:Goga_positionUpdate02} 
\end{eqnarray}
\end{subequations}
Combining the reformulated momentum update (eq.~\ref{eq:Goga_momentumUpdate}) and the two-step position update (eq.~\ref{eq:Goga_positionUpdate01} and \ref{eq:Goga_positionUpdate01}) yields the BAOA method
\begin{subequations}
\begin{align}
    p_k           &= p_{k- \frac{1}{2}}  -\nabla V(q_k) \Delta t  \label{eq:BAOA01} \\
    q_{k+\frac{1}{2}} &= q_k + p_k\frac{\Delta t}{2m} \label{eq:BAOA02} \\
    p_{k + \frac{1}{2}} &= + e^{-\xi \Delta t} p_k  + \sqrt{\left(1-e^{-2\xi \Delta t }\right)  m k_BT} \eta_k \label{eq:BAOA03} \\
     q_{k+1} &= q_{k+\frac{1}{2}} + p_{k+\frac{1}{2}}\frac{\Delta t}{2m}. \label{eq:BAOA04}
\end{align}
\end{subequations}
Following the nomenclature in Refs.~\onlinecite{Leimkuhler:2012} and \onlinecite{Leimkuhler:2013}, eq.~\ref{eq:BAOA01} is a $B$-step which updates the momenta according to the (deterministic) drift force, eqs.~\ref{eq:BAOA02} and \ref{eq:BAOA04} are two $A'$-half-steps which update the positions according to the current momenta, and eq.~\ref{eq:BAOA03} is an $O$-step which updates the momenta according to the stochastic force. 
Note that the $B$- and $O$-step use the full timestep $\Delta t$,
while the $A'$-update has been split into two half-steps, each with $\frac{\Delta t}{2}$.
We denote half-steps by $'$.
See SI, section I.A. for further details.

Since, eqs.~\ref{eq:Goga01}-\ref{eq:Goga04} and eqs.~\ref{eq:BAOA01}-\ref{eq:BAOA04} can be interconverted, the trajectories produced by these two algorithms for a given random number sequence $\boldsymbol{\eta}$ must be identical: $\boldsymbol{\omega}^\mathrm{GSD}=\boldsymbol{\omega}^\mathrm{BAOA}$.
This is the main result of this contribution.
%

\subsection{BAOA and BAOAB\label{sec:BAOA=BAOAB}}
We sketch the proof that BAOA\cite{BouRabee:2010} is equivalent to BAOAB \cite{Leimkuhler:2012,Leimkuhler:2013, Sivak:2013, Sivak:2014}. 
A similar argument can be made based on the propagators of the two algorithms\cite{Leimkuhler:2012}.
The BAOA algorithm iterates the following substeps:
$BA'OA'\,; BA'OA' \,; BA'OA'\,;  \dots$, where half-steps with $\frac{\Delta t}{2}$ are denoted by $'$ and semicolons show the end point of an integration cylce. 
A full $B$-step can be split into two consecutive $B'$-half-steps, which yields the following sequence of substeps:
$B'B'A'OA'\,; B'B'A'OA\,; 'B'B'A'OA'\,;   \dots$.
Since the algorithm is iterated, the choice of the endpoint of an integration cylce is somewhat arbitrary. One could denote the same sequence of steps as:
$B'\,; B'A'OA'B'\,; B'A'OA'B'\,; B'A'OA'  \dots$.
This is the BAOAB algorithm. (See SI section 1 for the algorithm and SI Fig.~1 for a side-by-side comparison of BAOAB and BAOA.)
There are two differences between the BAOA and the BAOAB algorithm. 
First, the initial state of the iteration differs by a $B'$-half-step.
To obtain identical intermediate results for BAOA and BAOAB for a given random number sequence $\boldsymbol{\eta}$, the initial momentum for BAOAB needs to be adjusted by a $B'$-half-step
\begin{eqnarray}
     p_0^{\mathrm{BAOAB}} = p_0^\mathrm{BAOA} - \frac{\Delta t}{2} \nabla V(q_0)  
\label{eq:adjustedInitialMomentum}
\end{eqnarray}
Second, the momenta, that are written to disc at the end point of an iteration, differ by a half-step $B'$.
In summary, (with adjusted initial momenta) GSD, BAOA and BAOAB write identical position trajectories to disc, but the momentum trajectory of BAOAB differs slightly from the GSD/BAOA momentum trajectory.
%

\section{Numerical results}

\subsection{Example trajectories \label{sec:example_paths}}

\begin{figure*}[!h]
    \centering
    \includegraphics[width=3in]{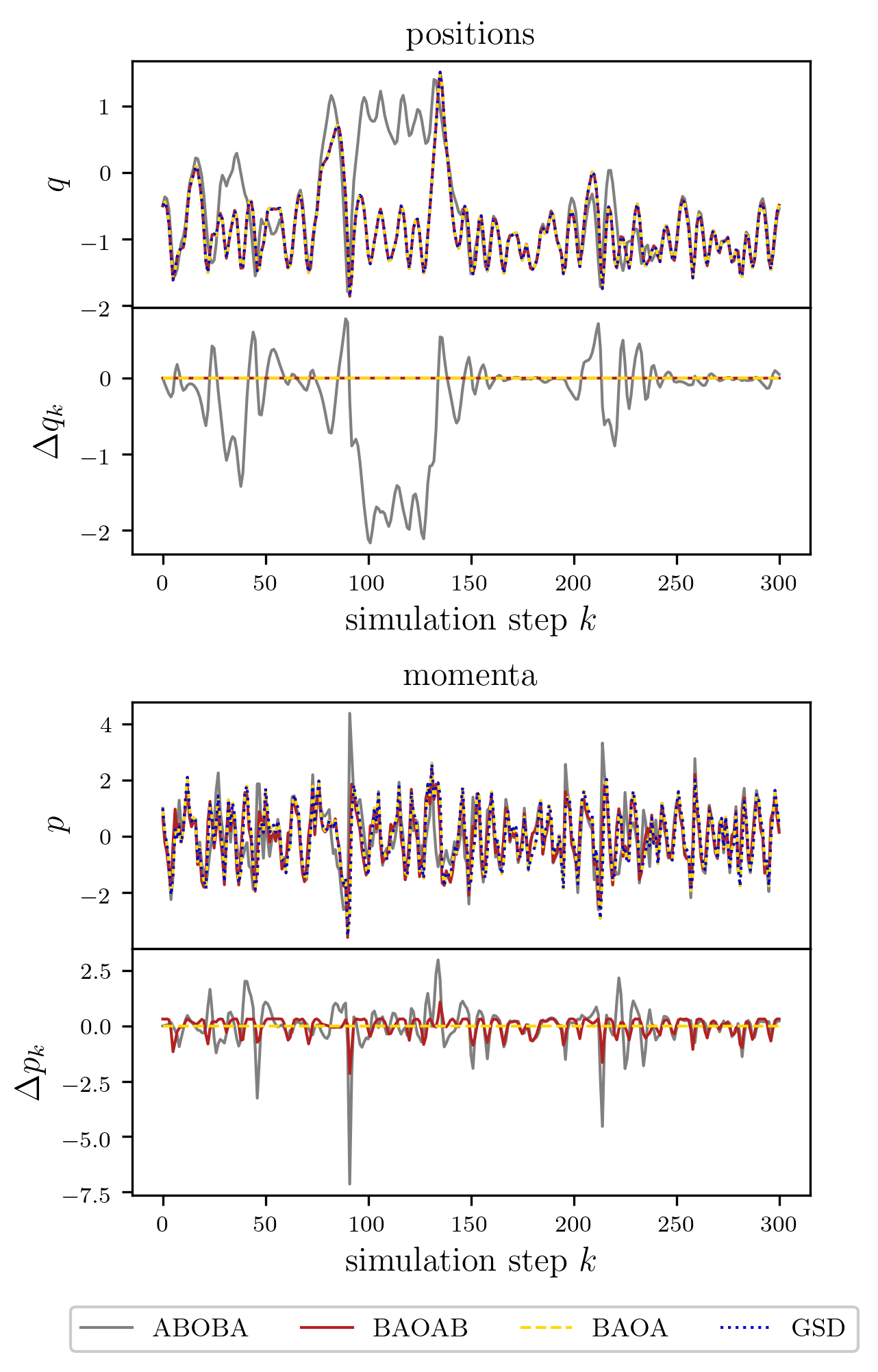}
    \caption{Example trajectories in 1D potential with timestep $\Delta t=0.25$.
    \textbf{Upper panel:} Position trajectories and deviations from GSD trajectory $\Delta q_k^I = q_k^{\mathrm{GSD}}-q_k^I$.
    \textbf{Lower panel:} Momentum trajectories and deviations from GSD trajectory $\Delta p_k^I = p_k^{\mathrm{GSD}}-p_k^I$.
    The initial momentum of BAOAB was shifted according to eq.~\ref{eq:adjustedInitialMomentum}.
    }
    \label{fig:ex_path_01}
\end{figure*}

To verify that GSD, BAOA and BAOAB yield the same position trajectories, we calculated example trajectories with these three algorithms. Additionally we included ABOBA \cite{Leimkuhler:2012, Leimkuhler:2013} in the comparison. (See SI section 1 for the algorithm.)
The trajectories were simulated in a one-dimensional tilted double well potential 
$V(q) = (q^2 - 1)^2 + q$.
%
%
To be able to compare the integrators, we generated a sequence of 300 normally distributed random numbers $\boldsymbol{\eta}$, and used this $\boldsymbol{\eta}$ for the integration with each of the integrators at $\Delta t=0.25$.
The initial conditions were $q_0=-0.5$ and $p_0=1$ for all integratores except BAOAB, for which the intial momentum was adjusted according to eq.~\ref{eq:adjustedInitialMomentum}.
(See SI section 2 for all simulation details.)
Fig.~\ref{fig:ex_path_01} shows the position trajectories and the momentum trajectories, as well as the deviations from the GSD trajectory: 
$\Delta q_k^I = q_k^{\mathrm{GSD}}-q_k^I$ and $\Delta p_k^I = p_k^{\mathrm{GSD}}-p_k^I$.
The position trajectories of GSD, BAOA and BAOAB are identical. 
Additionally, the momemtum trajectories of GSD and BAOA are identical, whereas the BAOAB momenta deviate slightly. 
This confirms our result from sections \ref{sec:GSD=BAOA} and \ref{sec:BAOA=BAOAB}.
The ABOBA trajectory deviates in both the positions and the momenta from the GSD/BAOA trajectory.
Note that for a small timestep, the differences between the trajectories of the four integrators are much smaller (SI Fig.~2).
This is expected, because all four integrators are guaranteed to converge to the true dynamics for $\Delta t \rightarrow 0$.
SI Fig.~3 shows the BAOAB trajectory for $\Delta t=0.25$, if the initial momentum is not adjusted.

\subsection{Numerical accuracy}

\begin{figure*}[!h]
    \centering
    \includegraphics[width=16cm]{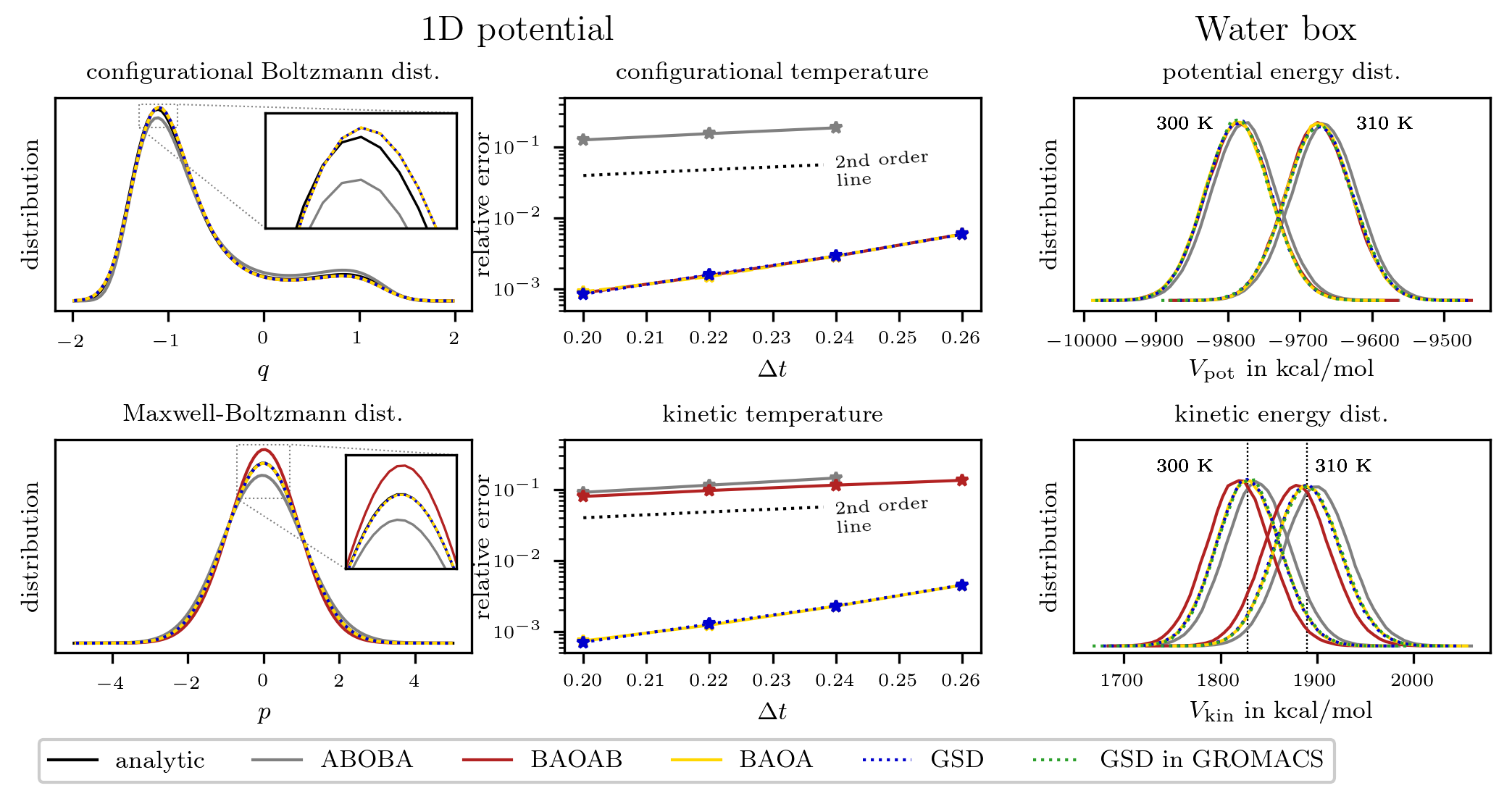}
    \caption{
    Numerical accuracy.
    \textbf{left column:} Configurational Boltzmann distributions with the inset magnifying the region around the deepest well (top) and Maxwell-Boltzmann distributions with the inset magnifying the region around the mean momentum (bottom) for the 1D potential and $\Delta t=0.25$. The analytic distributions are shown in black.
    \textbf{middle column:}
    Relative error in the average configurational temperature (top) and average kinetic temperature (bottom) for the 1D potential. The equation of the second order line is $\varepsilon^\mathrm{2nd}=\exp(\Delta t^2)$.
    \textbf{right column:}
    Distributions of the total potential energy (top) and total kinetic energy (bottom) for TIP3P bulk water at near-ambient conditions and temperatures 300 K and 310 K. 
    The vertical lines indicate the average kinetic energies expected by the equipartition theorem.
    ABOBA, BAOAB, BAOA and GSD were simulated with OpenMM. 
    The results for GSD in GROMACS were generated with GROMACS.
    Statistical uncertainties are smaller than the linewidth.
    }
    \label{fig:integrator_accuracy}
\end{figure*}

Next, we tested whether GSD and BAOA have the same numerical accuracy in the long-time limit. As model systems we use the one-dimensional tilted double-well potential from section \ref{sec:example_paths} and a box of 1024 TIP3P water molecules \cite{Jorgensen:1983}.
(See SI section 2.)

The left column in Fig.~\ref{fig:integrator_accuracy} tests how accurately ABOBA, BAOAB, GSD and BAOA reproduce the equilibrium distribution for the 1D potential in position and momentum space with a rather large timestep $\Delta t=0.25$. 
GSD, BAOA and BAOAB generate the same configurational distribution. 
Their distribution agrees well with the analytical Boltzmann distribution (SI eq.~4), whereas the distribution generated by ABOBA deviates from the analytical solution. 
GSD and BAOA yield the same very accurate momentum distribution.
(See SI eq.~5 for the analytical Maxwell-Boltzman distribution.)
BAOAB underestimates the variance in the momentum distribution leading to a higher peak at $p=0$. 
Conversely, ABOBA overestimates the variance leading to a lower peak at $p=0$.
These results confirm that BAOAB, GSD and BAOA perform equally in configurational sampling while BAOAB differs from GSD and BAOA in momentum space.
%

The middle column in Fig.~\ref{fig:integrator_accuracy} reports the accuracy with which the integrators reproduce the target temperature in the 1D potential. 
We show the relative error $\varepsilon(\Delta t) = (T_\mathrm{ref} - T_\mathrm{approx}(\Delta t))/T_\mathrm{ref}$ as function of the timestep $\Delta t$. 
Smaller relative errors imply better accuracy.
The relative error that is expected for an integrator with second order accuracy, 
$\varepsilon^\mathrm{2nd}=\exp(\Delta t^2)$, is shown by the dotted line \cite{Leimkuhler:2013}. 
The average temperature can either be computed as the configurational temperature $T_\mathrm{conf}$, 
which is an average with respect to the configurational Boltzmann distribution, 
or as the kinetic temperature $T_\mathrm{kin}$, 
which is an average with respect to the Maxwell-Boltzmann distribution
(see SI section 2.A).
In accordance with the observation that GSD/BAOA and BAOAB sample the same highly accurate configurational distribution, they also have the same low relative error in $T_\mathrm{conf}$. 
In comparison, ABOBA yields less accurate configurational temperatures with more than 10 \% discrepancy for all $\Delta t$.
For $T_\mathrm{kin}$ the accuracy of GSD/BAOA differs from the accuracy of BAOAB.
The relative error in the kinetic temperature sampled by GSD/BAOA is less than 1 \% error for any given timestep.
BAOAB yields less accurate kinetic temperatures with more than 10 \% discrepancy in the $\Delta t>0.23$ regime, similar to the accuracy of ABOBA.
This is line with the observation that GSD/BAOA samples the Maxwell-Boltzmann distribution very accurately. 

The right column in Fig.~\ref{fig:integrator_accuracy} shows an accuracy test for a molecular system. 
Following Ref.~\onlinecite{Rosta:2009} we determined the distributions of the total potential energy and total kinetic energy for bulk TIP3P water \cite{Jorgensen:1983} at near-ambient conditions and at two different temperatures, 300 K and 310 K, using a timestep of $\Delta t = 2$ fs.
As MD software package, we used OpenMM.
(See SI section II.B for all simulation details.)
We obtain an average potential energy of approximately -9780 kcal/mol at 300 K and -9670 kcal/mol at 310 K which is in good agreement with Ref.~\onlinecite{Rosta:2009}.
The potential energy distributions of GSD/BAOA and BAOAB are visually indistinguishable for both temperatures, whereas the distribution of ABOBA is shifted to slightly higher potential energies. 
GSD and BAOA yield kinetic energy distributions whose means agree very well with the average kinetic energy that is expected from the equipartition theorem (dotted lines). (See SI, section 2.B.)
The distribution of BAOAB is shifted to lower kinetic energies, which is in line with the underestimated variance of the Maxwell-Boltzmann distribution in the 1D-potential.
Conversely, the distribution of ABOBA is slightly shifted to higher kinetic energies.
To check the consistency across different MD software packages, we additionally simulated the TIP3P water box with GROMACS\cite{Abraham:2015}, i.e.~using GSD/BAOA.
The results are shown as a green dotted line in Fig.~\ref{fig:integrator_accuracy}. 
Both, the potential and the kinetic energy distribution, match the results of GSD/BAOA as implemented in OpenMM.


\subsection{Thermal rate constant}
\begin{figure*}[!h]
    \centering
    \includegraphics[width=8cm]{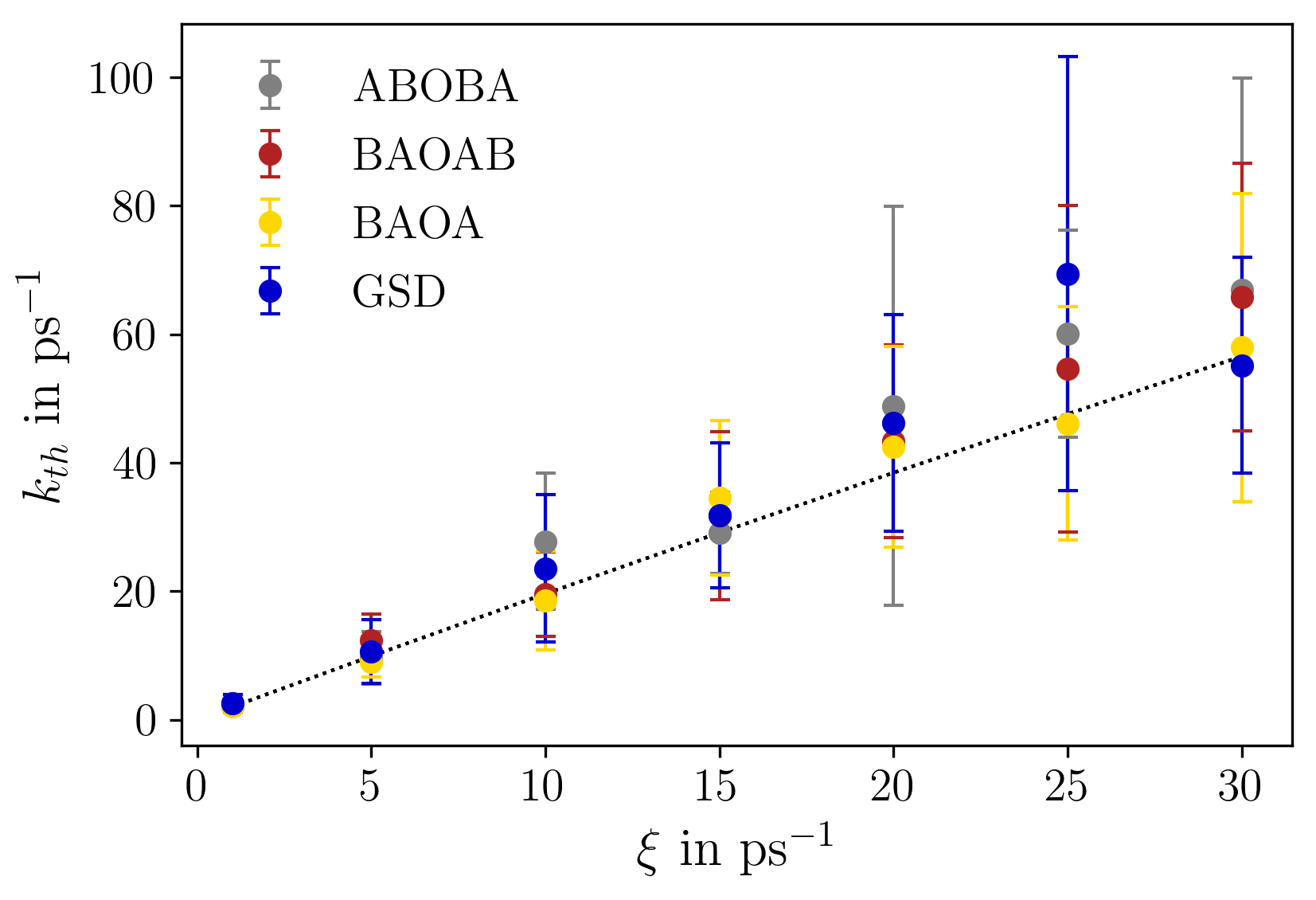}
    \caption{Mean and standard deviation of the thermal rate constants $k_{th}$ against the friction rate $\xi$ for the ideal gas model. The black line represents the analytic solution for an ideal gas with Langevin friction (eq.~\ref{eq:thermal_rate_ideal_gas}). }
    \label{fig:thermal_rates}
\end{figure*}
An important property of a Langevin integrator is its ability to provide a single exponential decay in temperature towards the target temperature after a temperature change, 
The decay rate is called thermal relaxation rate $k_{th}$ and can be adjusted via the friction rate $\xi$.
Increasing $\xi$ dampens the term related to the friction force $\exp(-\xi \Delta t) p$ and increase the associated random force term $\sqrt{(1-\exp(-2\xi \Delta t)m k_B T} \eta$ in the O-step of the Langevin integrators (e.g.~eq.~\ref{eq:BAOA03}). 
At high values of $\xi$, fewer simulation steps are necessary to drive the system towards the new target temperature, which corresponds to a high thermal relaxation rate $k_{th}$.
At low values of $\xi$, the response to a change in temperature is slow.
For an ideal gas with Langevin friction the relation between $k_{th}$ and friction rate $\xi$ is known analytically \cite{Goga:2012}
\begin{equation}
    k_{th}^\mathrm{analytic} = \frac{1 - \exp(-2 \xi \Delta t)}{\Delta t} \, .
    \label{eq:thermal_rate_ideal_gas}
\end{equation}
(dotted line in Fig.~\ref{fig:thermal_rates}).  
Following Ref.~\onlinecite{Goga:2012}, we determine the thermal relaxation rate $k_{th}$ from simulations of an  ideal gas in a cubic box for GSD, BAOA, BAOAB and ABOBA.
All four algorithms show the expected first-order decay towards the target temperature.
The thermal relaxation rates $k_{th}$ are in good agreement with the analytic solution for all four integrators (Fig.~\ref{fig:thermal_rates}). 
Their standard deviations increase with $\xi$, because at high friction rates the target temperature is reached within few timesteps ($\mathcal{O}(100\, \mathrm{fs})$) which generates a numerical uncertainty in the fit of the decay curve.


\section{Conclusion}

We have shown that GSD and BAOA are equivalent algorithms. 
Integrating eq.~\ref{eq:underdamped_Langevin_01} by either integrator yields the same position and momentum trajectory (for a given random number sequence $\boldsymbol{\eta}$). 
Consequently, GSD and BAOA  have the same numerical accuracy for both, configurational and kinetic properties.
BAOA samples the same positions as BAOAB (for a given random number sequence $\boldsymbol{\eta}$) \cite{Leimkuhler:2012, Song:2021}. Consequently, we can now state that GSD and BAOAB are equivalent configurational sampling algorithms.
Our simulations confirm that GSD/BAOA achieves the same high numerical accuracy as BAOAB for configurational properties.
We remark that other Langevin integrators have been compared to BAOAB and have also been found to exhibit similar or equivalent configurational properties \cite{Gronbech:2013, Sivak:2014, Li:2017, Zhang:2019, Finkelstein:2021}.
GSD/BAOA and BAOAB sample slightly different momentum trajectories.
%
We find that GSD/BAOA samples the marginal distribution of momenta more accurately than BAOAB, thus achieving excellent accuracy for configurational as well as kinetic properties.
A similar finding is mentioned in the documentations of the MD package OpenMM \cite{OpenMMDoc} and the MD package AMBER 2021 \cite{AMBERDoc}, where GSD/BAOA is called LFMiddle integrator (see eqs.~16 and 17 in Ref.~\onlinecite{Zhang:2019}) as well as in Ref.~\onlinecite{Zhang:2019}.
This finding is surprising, because BAOAB, being a symmetric Langevin integrator \cite{Leimkuhler:2016c}, would be expected to show better convergence than a non-symmetric Langevin integrator.

Our results have several practical implications. 
Any analysis or benchmark of the configurational accuracy obtained for one of the three integrators equally applies to the other two integrators. 
In particular, the stability of BAOAB with respect to large timesteps also applies to GSD/BAOA.
Similarly, any extension of one of the integrators (e.g. incorporation of constraints \cite{Peters:2014, Leimkuhler:2016b, Zhang:2019} or multi-timestep algorithms \cite{Leimkuhler:2013b, Lagardere:2019}) can straightforwardly be transferred to the other two.
The results also help in the development of path reweighting methods \cite{Donati:2017, Donati:2018, Kieninger:2021}, because equal Langevin integrators have equal path reweighting factors.
Finally, with the combined experience gained with the GSD/BAOA and the BAOAB integrator, the MD community has a configurational sampling algorithm that has been tested and proven robust for system sizes and particle resolutions ranging from dissipative particle dynamics over coarse-grained and atomistic MD to path-integral MD.
%

\section*{Supporting information}
Details on the computational methods and additional numerical results are reported in the supporting information. All input files and scripts are provided on our GitHub repository \cite{GitHub}.

\section*{Acknowledgements}
We would like to thank Alexander H.~de Vries, Nicu N.~Goga, Benedict Leimkuhler and Gabriel Stoltz and both reviewers for insightful discussions.
This work was funded by Deutsche Forschungsgemeinschaft (DFG): project ID 235221301 (CRC 1114), project ID 431232613 (SFB 1449) and under Germany´s Excellence Strategy – EXC 2008/1 – 390540038.

\bibliography{literature}



%

%

\end{document}